\documentclass[12pt,preprint]{aastex}

\def\sqig{$\sim$}
\def\sun{$_\odot$}
\def\degrees{$^{\circ}$}

\def\source{XTE\,J1855-026}
\def\src{XTE\,J1855-026}

\begin{document}
\title{
The Orbit and Position of the X-ray Pulsar XTE J1855-026
- an Eclipsing Supergiant System}

\author{Robin H.D. Corbet\altaffilmark{1} and Koji Mukai\altaffilmark{1}}

\affil{Laboratory for High Energy Astrophysics, Code 662,\\
NASA/Goddard Space Flight Center, Greenbelt, MD 20771}
\altaffiltext{1}{Universities Space Research Association}
\email{corbet@gsfc.nasa.gov}

\begin{abstract}
A pulse timing orbit has been obtained for the X-ray binary
\src\ using observations made with the Proportional Counter 
Array on board the Rossi X-ray Timing Explorer.
The mass function obtained of \sqig 16M\sun\ together
with the detection of an extended near-total eclipse
confirm that the primary
star is a supergiant as predicted. The orbital eccentricity is found to be
very low with a best fit value of 0.04 $\pm$\ 0.02. The orbital period is also refined
to be 6.0724 $\pm$ 0.0009 days using
an improved and extended light curve
obtained with RXTE's All Sky Monitor. Observations with
the ASCA satellite provide an improved source
location of R.A. = 18$^{h}$ 55$^{m}$ 31.3$^{s}$, 
decl. = -02\degrees\ 36\arcmin\ 24.0\arcsec\ (2000)
with an estimated systematic uncertainty of less than 12\arcsec.
A serendipitous new source, AX J1855.4-0232, was also discovered
during the ASCA observations.

\end{abstract}

\section{Introduction}

The X-ray source, \src, was discovered during Rossi
X-ray Timing Explorer (RXTE) scans along the
galactic plane  (Corbet et al. 1999; hereafter Paper I) .
The source showed pulsations at a period of 361 s and a light
curve obtained with RXTE's All Sky Monitor (ASM) showed
modulation at a period of 6.067$\pm$ 0.004 days which was interpreted as the
orbital period of the system. The X-ray spectrum above \sqig3 keV could
be fitted with an absorbed power law model with a high-energy cut-off,
and an iron emission line at approximately 6.4 keV. These
results, in particular the location of the source
in the orbital period/spin period diagram (Corbet 1986), were interpreted 
as indicating that \src\ is likely to consist of a neutron star
accreting from the wind of an O or B supergiant primary. A less likely
interpretation was that \src\ is instead a Be/neutron star binary,
in which case it would have an unusually short orbital period for such
a system.

Here we present the results of observations made with the RXTE Proportional
Counter Array (PCA)
that were performed over the course of one complete orbital
cycle. The observations were performed with the aims of (i) measuring
the orbital parameters to determine the X-ray mass function and thus
the nature of the primary star, and (ii) determining whether a eclipse is
present in the light curve. 
A system containing a supergiant primary rather than a main-sequence
Be star would be much more likely to exhibit an eclipse due to
the much greater size of the primary star.
We also report on observations made with the imaging detectors onboard
the ASCA satellite which enable the source position to be refined.
RXTE ASM observations utilizing this improved position
and extending over six years allow further refinement
of the orbital period.
Spectroscopic results from both satellites are not
discussed here and will be presented elsewhere.

\section{Observations}

\subsection{RXTE Observations}

In this paper we present the results of observations of \src\ that have
been made with two of the instruments on board RXTE (Bradt, Rothschild, \&
Swank 1983): the All
Sky-Monitor (ASM) and the Proportional Counter Array (PCA).

The ASM (Levine et al. 1996) consists of three similar
Scanning Shadow Cameras, sensitive to X-rays in an energy band of
approximately 2-12 keV, which perform sets of 90 second pointed
observations (``dwells'') so as to cover \sqig80\% of the sky every
\sqig90 minutes.  
The Crab produces approximately 75
counts/s in the ASM over the entire energy range. Observations
of blank field regions away from the Galactic center suggest that
background subtraction may produce a systematic uncertainty of about 0.1
counts/s (Remillard \& Levine 1997). The ASM light curve of \src\
considered here now covers approximately 6 years compared to the
less than 3 years reported in Paper I. In addition, the light curve
is improved because a new ASM light curve was generated using
the improved source position that we determine with ASCA (Section 3.2).
Because ASM fluxes are determined using a model fitting procedure
which uses cataloged source locations, uncertainties in a source's position
can lead to larger uncertainties in the measured X-ray fluxes. 
The ASM data were further filtered by excluding all dwells
where the modeled background in the lowest energy band was
greater than 10 counts/s. This procedure helps to exclude points
where the data are contaminated by solar X-rays.

The PCA is described in detail by Jahoda et al. (1996).  This detector
consists of five, nearly identical, Proportional Counter
Units (PCUs) sensitive to X-rays with energies between 2 - 60 keV with
a total effective area of \sqig6500 cm$^2$. The PCUs each have a
multi-anode xenon-filled volume, with a front propane volume which is
primarily used for background rejection.  
The Crab produces 13,000
counts/s for the entire PCA across the complete energy band.  The PCA spectral resolution at 6 keV is approximately 18\%
and the field of view is 1\degrees\ full width half maximum (FWHM).
PCA observations of \src\ were obtained over the course of
one complete binary orbit in November 1999. Observations
were not continuous due to both interruptions because
of instrumental constraints such as Earth occultations of
the source and passages through the South Atlantic Anomaly when
the instruments are not operated, and because observations of
other sources were also undertaken during this time. The
resulting total exposure time was 150 ks. Due to instrumental
problems not all of the PCUs are always operated during an
observation and the typical number of PCUs turned on at any one time was three.
To give consistency in the analysis presented here we make use of
data collected by the two PCUs (numbers 0 and 2) which were always
operated. Data extraction, including background subtraction, followed
standard procedures. The light curve used in the analysis presented
here includes photons in the energy range of approximately 2.5 to 24 keV.

In addition to the ASM and PCA, RXTE also carries 
the HEXTE experiment (Rothschild et al. 1998)
which is sensitive to high energy X-rays in the
range of 15 to 250 keV. However, results from this
experiment are not presented here as its smaller collecting
area, together with the lower source photon flux at
higher energies, make HEXTE less useful for pulse timing. HEXTE data
will instead be presented together the with PCA and ASCA spectral
results.


\subsection{ASCA Observations}

Observations of \src\ were made with the ASCA X-ray astronomy satellite
(Tanaka, Inoue, \& Holt 1994)
on 1999 October 14 from 00:21:38 to
11:52:43 and from October 15 15:34:01 to October 15 03:42:01.
These observations 
had durations of 40.3 and 41.5 ks respectively
and were timed to coincide with the
predicted times of orbital minimum and maximum X-ray flux.
ASCA carried four sets of X-ray telescopes, two of which were equipped
with solid-state SIS detectors (Burke et al. 1993) and two with
Gas Imaging Spectrometer detectors (GIS, Makishima et al. 1996,
Ohashi et al. 1996). The intrinsic point spread function of the ASCA
mirrors themselves had a core of FWHM ~50\arcsec\
(Jalota, Gotthelf, \& Zoonematkermani 1993). This was greatly
oversampled by the SISs but the GISs' own spatial resolution is
comparable to the mirror point spread function. For bright sources,
positions can be determined to an accuracy of 12\arcsec\ radius
when temperature dependent errors in the attitude 
solution are compensated for (Gotthelf et al. 2000).

\section{Results}
\subsection{RXTE}
The additional data
obtained from the ASM over the more than three years period since the
results presented in Corbet et al. (1999) and the improved
flux measurement accuracies were used to determine a
more precise value for the orbital period. From fitting a sine
wave to the orbital modulation a period of 6.0724 $\pm$ 0.0009 days is obtained.
While this 0.001 day error is the statistical 1$\sigma$ error from the $\chi^2$ fitting,
it is possible that systematic effects may make the measurement somewhat less
accurate than this if, for example, there are 
systematic effects such as changes in the orbital
modulation that are not independent from cycle to cycle.
The ASM light curve folded on this period is shown in Figure 1.

The light curve obtained with the PCA is shown in Fig. 2. A clear
feature of this light curve is a near total eclipse.
During this period the X-ray flux is 1.03 $\pm$ 0.01 (statistical) counts/s/PCU.
Although eclipse ingress and egress were not observed we
can use our light curve to derive limits on the length of the
eclipse of 1.63 $>$ T $>$ 0.42 days. 
Examination of shorter stretches of the data shows that pulsations
are present at all times except during the eclipse.  Bright flares can
be seen at approximately 3.5$\times$10$^5$ s after
the start of the observation.
When these flares are examined in detail (Fig. 3)
the flaring maxima appear to occur at the maximum of the pulse profile.

In order to determine the orbital parameters from the pulse arrival
times a reiterative procedure was used. An initial pulse template was
constructed by folding the PCA light curve on a value for the pulse period
obtained from a power spectrum of the entire light curve. The profile
was binned into 720 bins thus giving approximately 0.5s time resolution.
The light curve was then divided into sections using individual
``good time intervals'' (i.e. continuous data stretches between
Earth occultations, passages through the high particle background
regions where the detectors were turned off, and
observations of other sources). For each of these intervals
a pulse profile was constructed by folding on the estimated period
and the relative phase compared to the template was calculated by
cross-correlation. An initial circular orbit was then calculated
from the relative phase changes. The light curve was then corrected
for this orbit and a new sharper pulse template derived.
This procedure was performed several times until no significant change
in orbital parameters occurred on the next iteration. 
The resulting mean pulse profile is shown in Fig. 4. This shows
that, in addition to an overall quasi-sinusoidal modulation, there are
also sharp features in the pulse profile. As long as they are 
always present,
such features aid in obtaining greater precision in the pulse delay
curve.

The pulse delay curve is shown in Fig. 5 together with a circular
orbit fit. 
It can be seen that a small number of the pulse delay values (three)
show much larger deviations from the orbital fit than do the other points.
These points were investigated in more detail but do not show
any large difference from the other data points. They have maximum values
of the cross-correlation function that are not especially small, and
the X-ray flux is not significantly lower or higher than for neighboring
data stretches. It is possible that some type of small flare occurred
that was not modulated with the usual pulse profile. However,
the very large flares that can clearly be seen in the light curve
were {\em not} accompanied by any anomalous pulse arrival time
effects and, for those flares, the usual pulse modulation continued
throughout the flares. In Fig. 6 we show all the individual profiles.
While pulse profile changes can be seen, the discrepant points
do not appear to show obvious peculiarities.
Performing orbital fits both with and without these
discrepant points does not greatly change the best fit values.
However, it does result in much larger errors on the parameters
if error bars on individual points 
are scaled so as to give a reduced $\chi^2$ of 1.
The results from fits using both edited and unedited data sets
and employing both circular and eccentric orbit models
are given in Table 1.
As there is no significant detection of an orbital eccentricity
(e = 0.04 $\pm$ 0.02 for the edited data set)
in Figures 1, 2 and 5
where orbital phase is marked we use the circular orbit together
with T0 defined as the predicted eclipse center.
The resulting pulse period of 360.741 $\pm$ 0.002 s is consistent with the
value of 361.1 $\pm$ 0.4 s
found in Paper I.
In order to determine the limits on pulse period changes a two
step process was necessary because the data only span a single orbit.
For all four cases the fits were initially done with $\dot{P}$ fixed
at zero. Next, $a\ sin\ i$ was fixed at the value found
from the first fit, and $\dot{P}$ was allowed to vary. 
This process is similar to that used by Clark (2000).
Using the circular orbit ephemeris we find the PCA limits on the times of eclipse
ingress and egress to be: $-0.138 < \phi_{ingress} <  -0.033$,
and $0.036 < \phi_{egress} < 0.131$. 

\subsection{ASCA}
From the ASCA observations at orbital maximum a position is found of
R.A.= $18^h55^m31.3^s$, decl.= $-02$\degrees\ 36\arcmin\ 24.0\arcsec\ (2000) in SIS-0 and SIS-1, after
correcting for the temperature-dependent attitude solution error
(Gotthelf et al. 2000) present in the current (revision 2) processing.
These coordinates are statistically accurate to about 2.5 SIS pixels
(4\arcsec).  The dominant error, however,  is the residual systematic uncertainty
in the attitude reconstruction, estimated to be 12" (90\% confidence;
Gotthelf et al. 2000).  The source position as determined from the GIS-2 data,
although subject to a larger uncertainty, is consistent with the SIS
position.  The GIS-3 data are unsuitable for this purpose, as the source
was observed very near a window support mesh on this instrument.
A finding chart based on this position is
shown in Fig. 7 using a red Space Telescope
Science Institute Digitized Sky Survey image.
While there is no obvious bright candidate within the error circle
we note the presence of an object at the NW of the region.

Our new orbital ephemeris shows that the two ASCA observations were
obtained during orbital phase ranges of 0.020 - 0.099 (orbital minimum) and 
0.288 - 0.372 (orbital maximum). There is no sign of an eclipse egress
in the ASCA observations at orbital minimum thus extending the
limit on eclipse egress beyond the PCA limit of $\phi_{egress} >$ 0.036
to give combined limits of $0.099 < \phi_{egress} < 0.131$.
The combined PCA and ASCA limits on eclipse ingress and egress
are marked on the folded ASM light curve in Figure 1.

In addition to \src, another faint source is detected in the ASCA
observations and is most prominent in the observations
obtained during the eclipse of \src.
The coordinates of this new source are found to be
R.A. = $18^h 55^m 28.0^s$, decl. = $-02$\degrees\ 32\arcmin\ 33\arcsec\
(2000) and the flux is 4 $\pm$ 1 $\times$ 10$^{-13}$ ergs cm$^{-2}$
s$^{-1}$ (1 - 5 keV).  Due to the faintness of this new source (AX
J1855.4-0232) the position uncertainty is greater at approximately
1\arcmin.

\section{Discussion}

The small residual flux seen by the ASM
during the eclipse of \src\ (Figure 1) may be interpreted as
a systematic measurement error. The residual PCA flux of approximately
1 count/s/PCU corresponds
to only \sqig 0.03 ASM counts/s. This indicates that there is
a systematic offset in the ASM fluxes for \src\ of about 0.15 ASM counts/s
which is consistent with the expected accuracy.
A contribution to the PCA and ASM fluxes during eclipse will also occur
from AX J1855.4-0232 which is only \sqig4\arcmin\ from \src\ and thus not
far from
the peak of the 1\degrees\ FWHM PCA collimator response.
Evidence for the presence of the eclipse can also be seen in the folded
ASM curve but it is difficult to extract eclipse constraints from this
because of the low source count rate in the ASM.

The discrepant points in the pulse delay curve may perhaps be
caused by brief changes in the pulse profile. However, if this
occurred it might be expected to be related to overall flux which
does not appear to be the case.
We note the presence of a flare at an orbital phase of about 0.5.
Although there appears to be no known reason to expect flares at this phase
in particular
we note that two other supergiant systems have light curves that
have exhibited flaring near this orbital phase.  These are 2S\thinspace 0114+650
(Hall et al. 2000) and X\thinspace 1538-522 (Corbet et al. 1993). However, this
small number of examples does not yet give definite evidence that this
is a real phase related effect.

If we assume that the eclipse is symmetric around phase 0 for
the circular orbit fit then we can obtain somewhat stricter limits
on the duration of the eclipse. With this assumption, and
the combined ASCA and PCA results, we thus find that the total
phase duration of the eclipse is in the range
of 0.198 to 0.262,
and the corresponding angular half width is 36\degrees $< \theta_e <$ 47\degrees. The lower limit on the eclipse duration implies a minimum
radius of the mass donating star of approximately 50 lt-s or 20 R\sun.
This radius is consistent with that of a B0I star, comparable
with the primaries in other wind-accretion driven high-mass
X-ray binaries (Liu, van Paradijs \& van den Heuvel 2000). 
While the orbital period and pulse period are also comparable with
parameters measured for similar systems the low orbital eccentricity
we find is apparently the lowest known
for this class (Bildsten et al. 1997, Clark 2000).

\section{Conclusion}

The light curve and pulse timing orbit clearly show \src\ to be a
supergiant X-ray binary as predicted in Paper I. With
the detection of the eclipse and timing measurements over an entire
orbit
the system parameters can now be determined. Future pulse timing
observations would enable a search for orbital period
changes as seen in some other high mass X-ray binaries (e.g.
Clark 2000, Levine, Rappaport, \& Zojcheski 2000, and references therein).

If an optical or IR counterpart could be found and its radial
velocity orbit measured this would be valuable as the
system would then be a ``double-lined" eclipsing binary
and the neutron star mass could be directly determined. 
While the optical reddening to this
object implied by the measured X-ray absorption is high
(N$_H$ = 15$\times$10$^{22}$ cm$^{-2}$ $\Rightarrow$ E(B-V) = 24, Paper I)
at least
some of this absorption may be local to the X-ray source rather than
genuinely interstellar. 
Tighter constraints on
the eclipse duration will also be valuable in obtaining precise measurements
of the system parameters.

\acknowledgments
We thank R.A. Remillard
for producing the revised ASM light curve.
This paper made use of the Digitized Sky Survey produced at
the Space Telescope Science Institute.

\pagebreak
\noindent
{\large\bf Figure Captions}

\figcaption[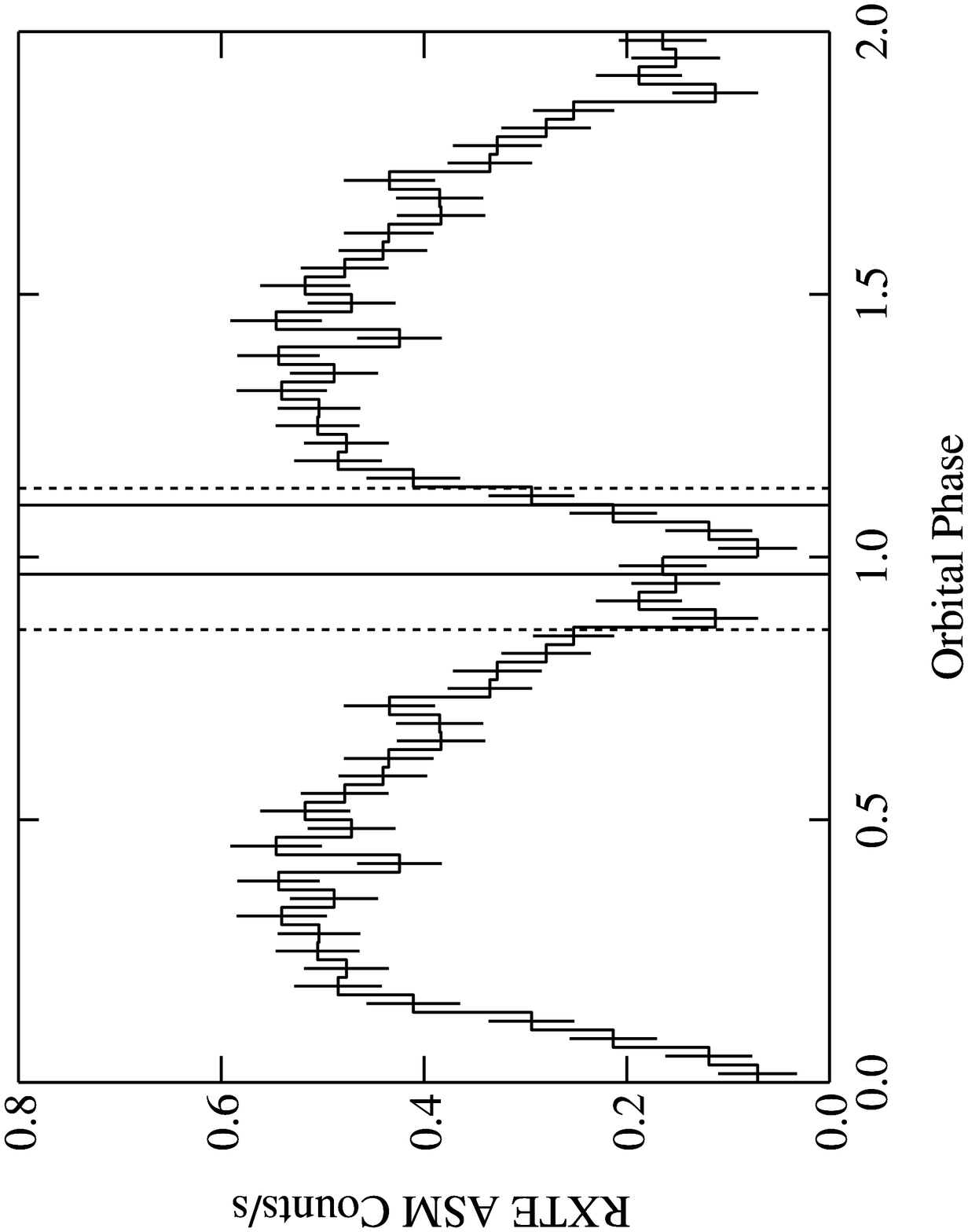]{The RXTE ASM light curve of \src\ folded
on the orbital period. The solid and dashed vertical lines indicate
the lower and upper limits on the eclipse duration respectively
as derived from the PCA and ASCA observations.}

\figcaption[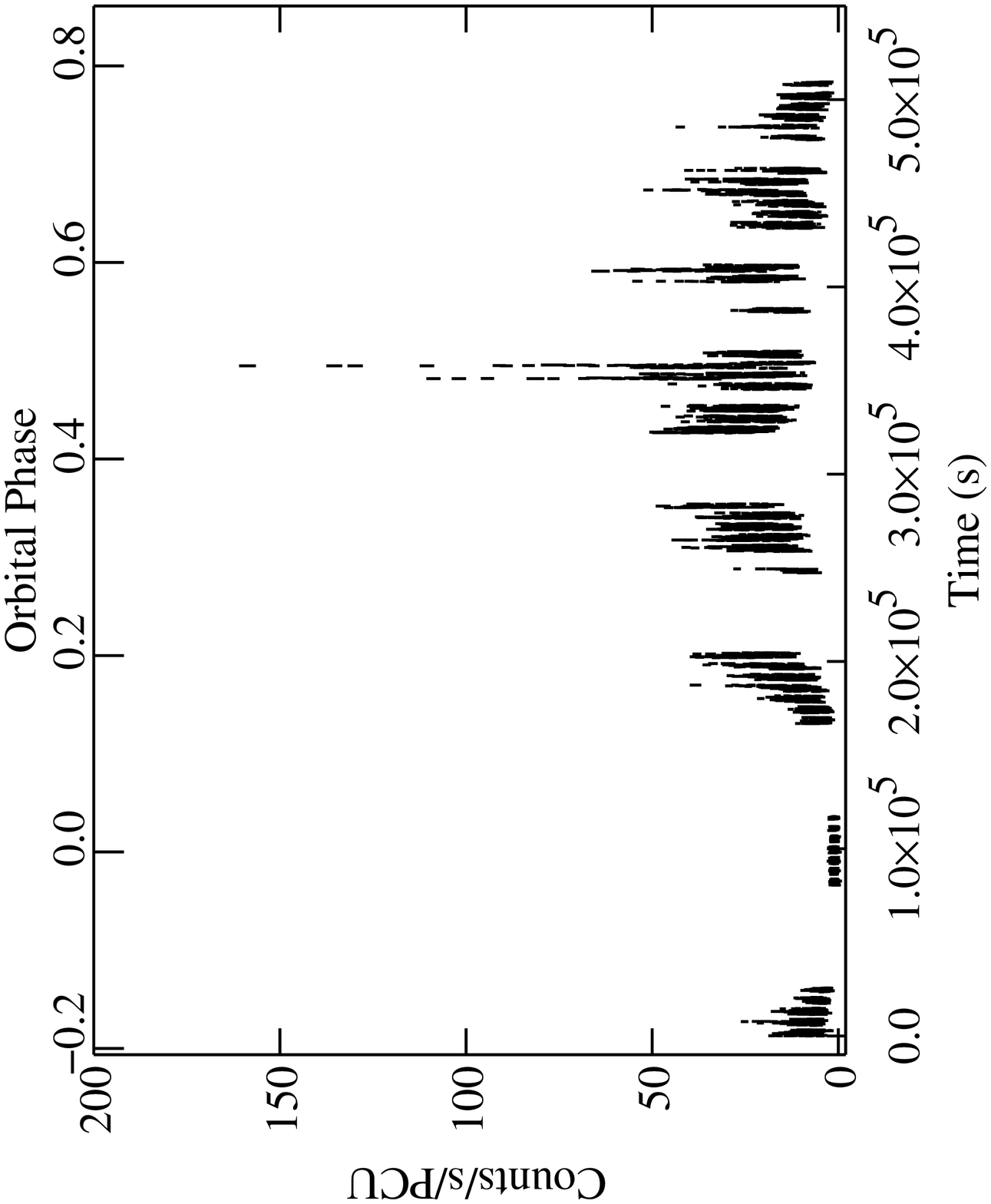]{The background subtracted light curve of \src\ obtained
with the RXTE PCA. Time is relative to the start of the
observation at MJD 51488.066}

\figcaption[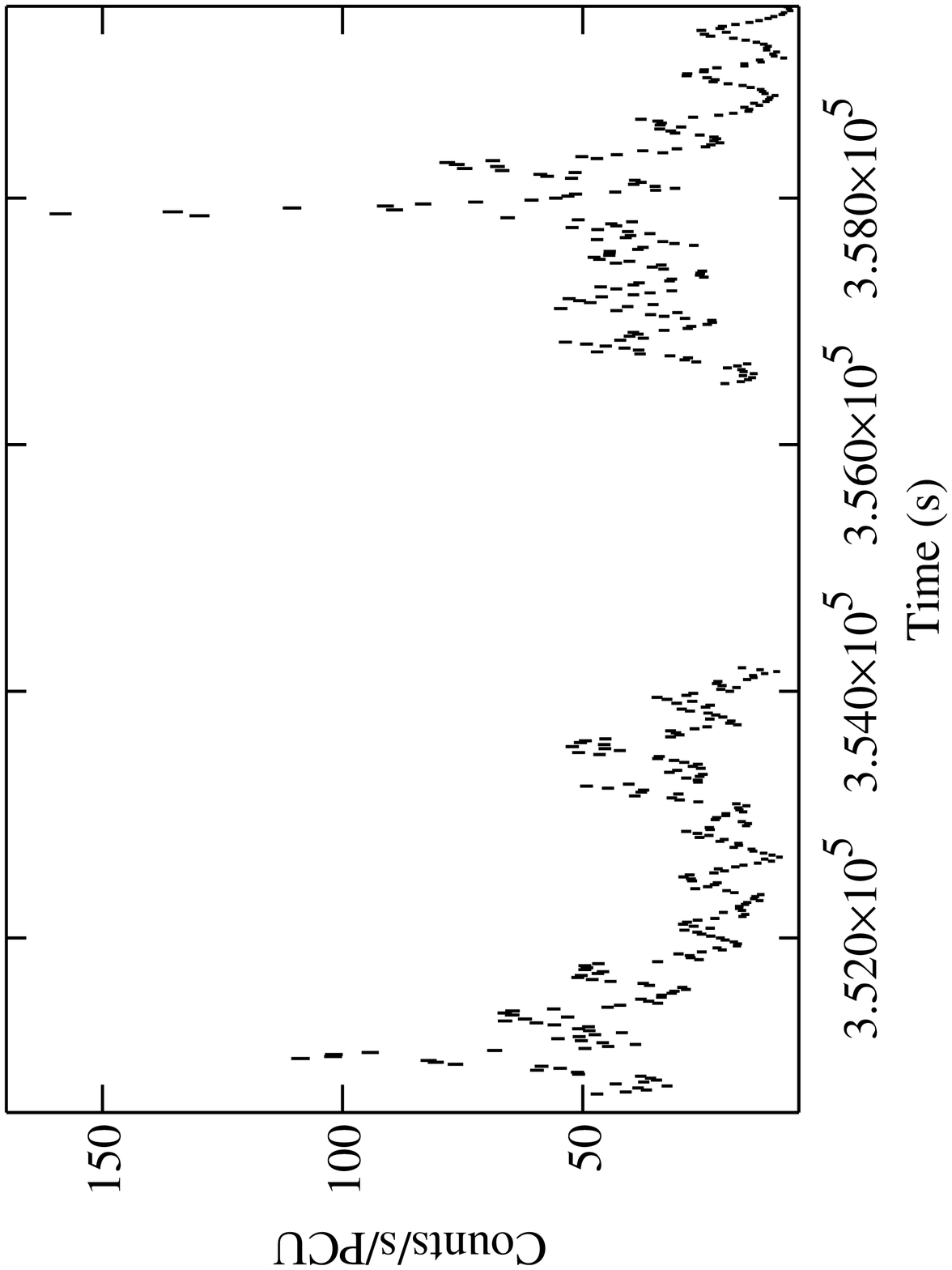]{Detail of the background subtracted light curve
of \src\ showing the two flares.}

\figcaption[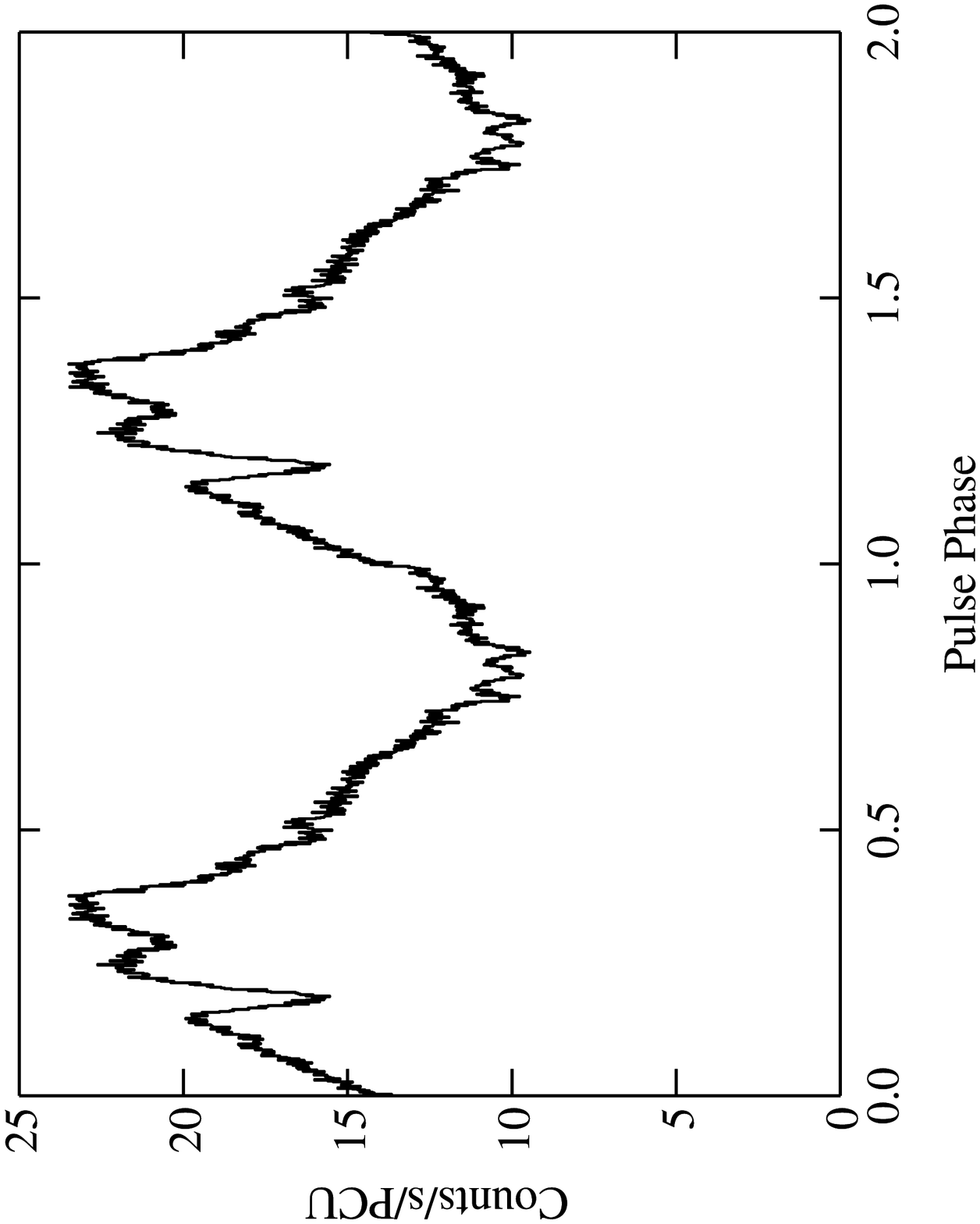]{The mean background subtracted pulse profile of \src\ obtained
with the RXTE PCA}

\figcaption[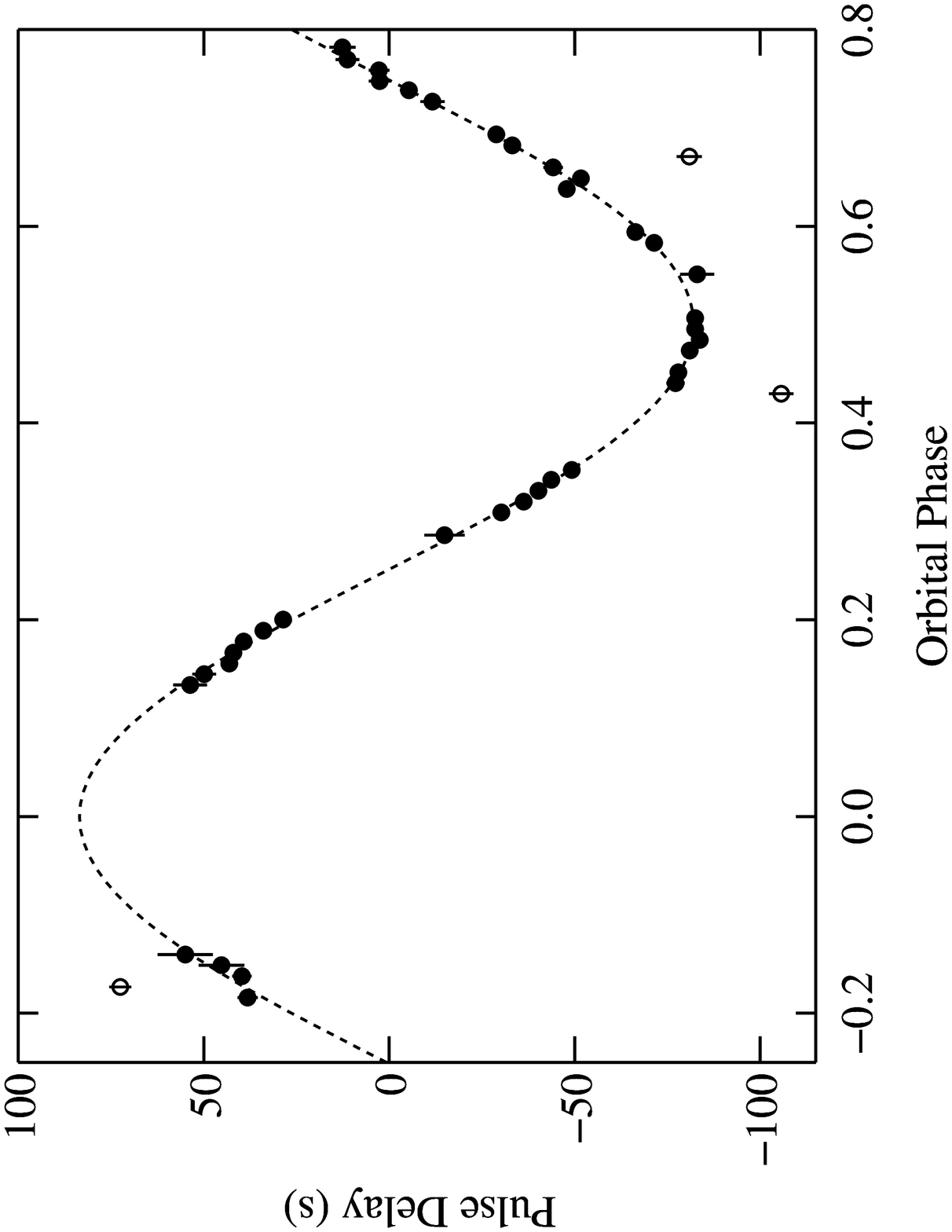]{The pulse delay curve for \src. The dashed line indicates
the best fit circular orbit. Three data points with values very discrepant
from the curve, marked as open circles in the plot,
were excluded from the fit.}

\figcaption[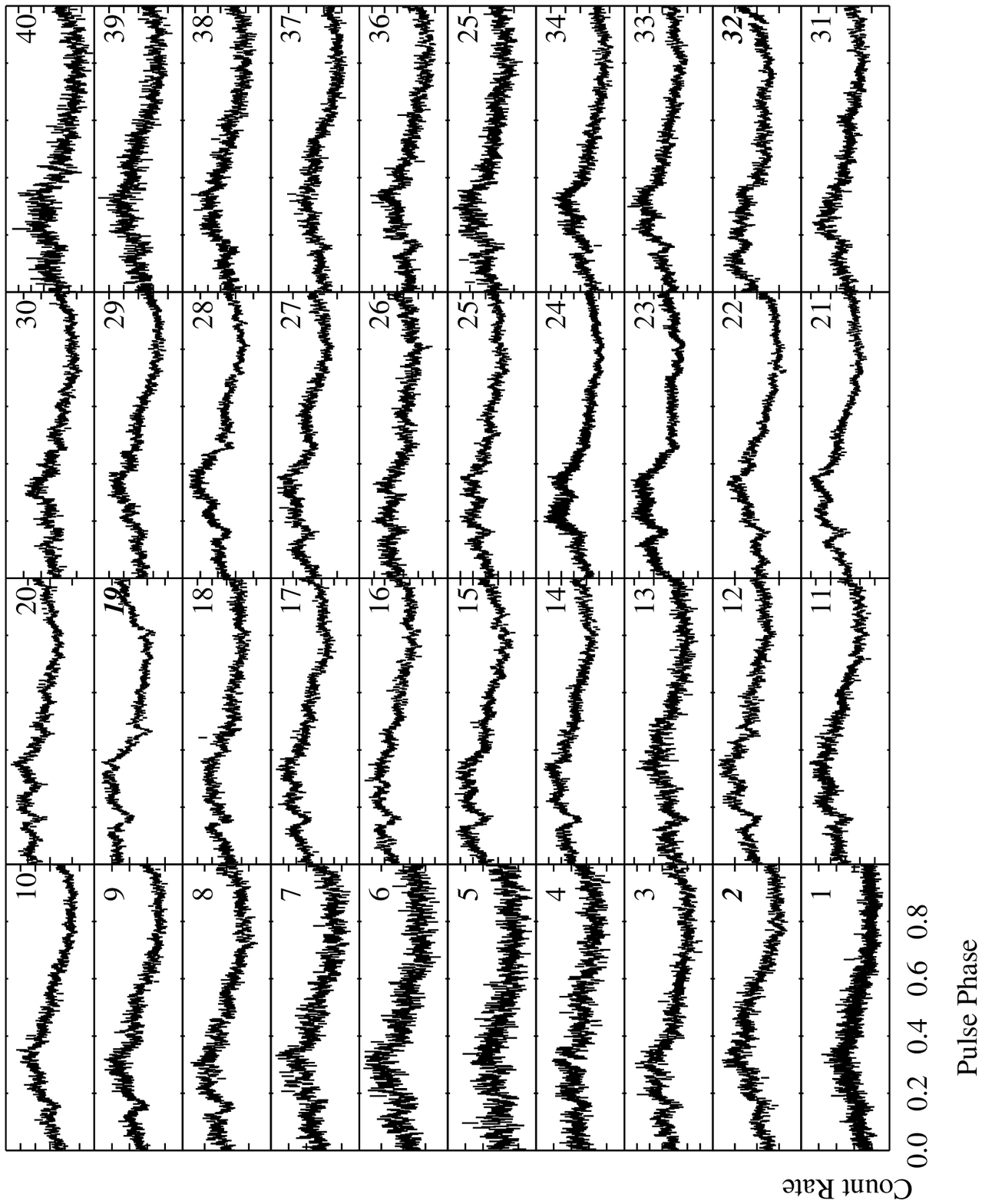]{The pulse profiles of \src\ used in the pulse
arrival time analysis. Pulses numbers 2, 19, and 32 have exceptionally
large discrepancies from the fitted orbit.}

\figcaption[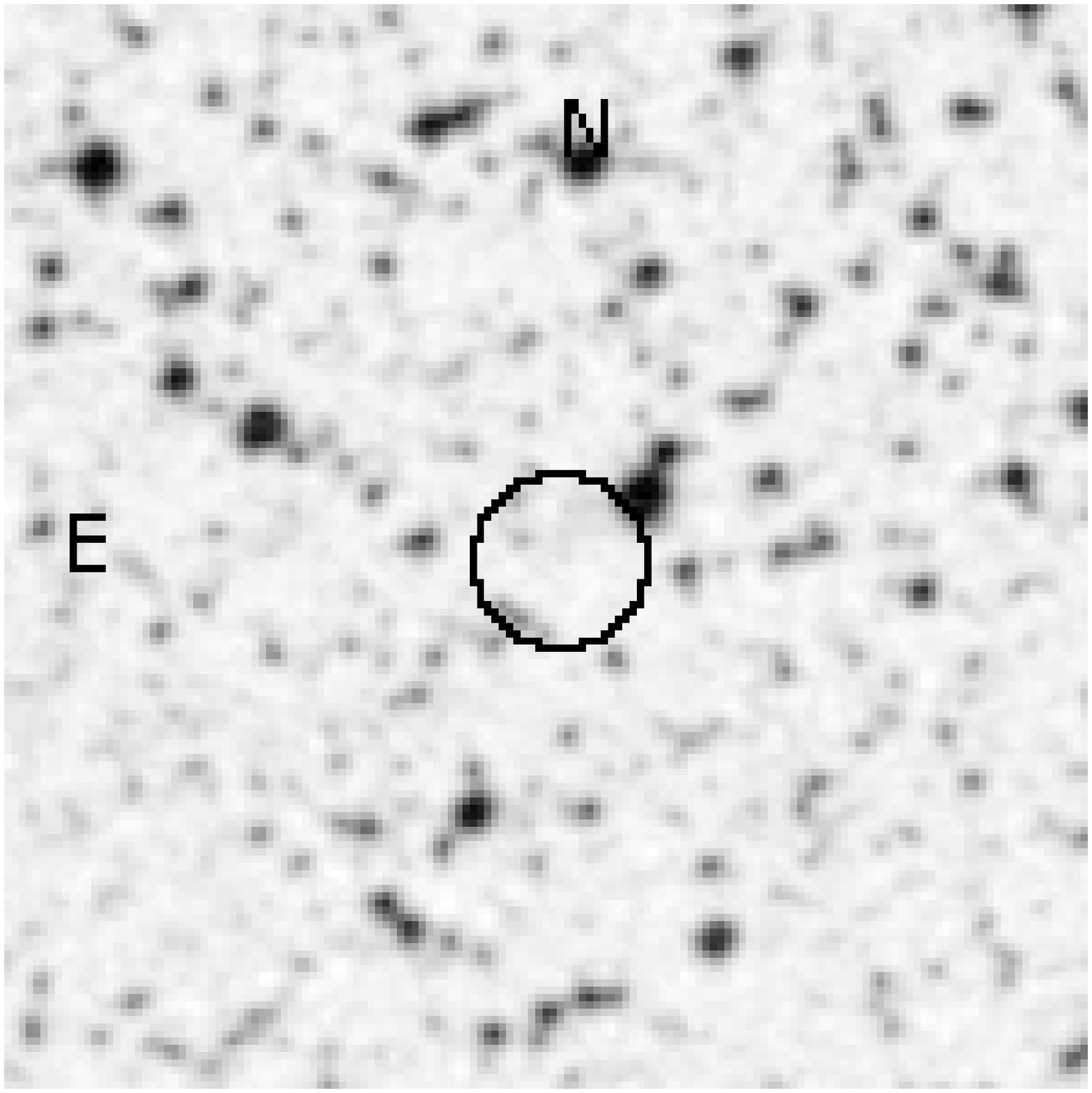]{A Space Telescope Science Institute Digitized Sky
Survey (red, second generation) image centered on the position of \src\ found from the
ASCA observations.  North is at the top and East at the left.  Image size
is 2.5\arcmin $\times$ 2.5\arcmin.}

\begin{table}
\caption{Orbital Parameters of \src}
\begin{center}
\begin{tabular}{lcccc}
          & All Data  &  All Data &  Edited Data & Edited Data\\
Parameter & Eccentric &  Circular &  Eccentric   & Circular\\
\tableline
$P_{pulse}$ (s)  & 360.733 $\pm$ 0.006 & 360.734 $\pm$ 0.005 & 360.739 $\pm$ 0.002 &  360.741 $\pm$ 0.002 \\
$\dot{P}_{pulse}$ (s s$^{-1}$ $\times$ 10$^{-8}$) & 3.7 $\pm$ 12 & 2.1 $\pm$ 11 & 1.7 $\pm$ 4.4 & 1.5 $\pm$ 3.6\\
a sin i (lt s)   & 81.8 $\pm$ 4.2 & 82.4 $\pm$ 2.4 & 80.5 $\pm$ 1.4 & 82.8 $\pm$ 0.8\\
T0 (MJD - 51400)) &  Undefined & 95.25 $\pm$ 0.02 & 91.4 $\pm$ 0.3 & 95.276 $\pm$ 0.007 \\
e & 0.0 $\pm$ 0.06 & -- & 0.04 $\pm$ 0.02 & -- \\
$\omega$ (degrees) & Undefined & -- & 226 $\pm$ 15 & -- \\
$\chi^{2}_{\nu}$ (d.o.f.) & 11.5 (34) & 10.8 (36) & 1.0 (31) & 1.06 (33) \\
\tableline
Orbital period (days) & 6.0724 $\pm$ 0.0009 & -- & -- & -- \\
Mass function (M\sun) & 15.9 $\pm$ 2.5 & 16.3 $\pm$ 1.4 & 15.2 $\pm$ 0.8 & 16.5 $\pm$ 0.5\\
\tableline 
\end{tabular}
\end{center}
All parameter errors are 1$\sigma$ single-parameter confidence levels.
Errors on individual pulse timing measurements were scaled to make
$\chi^{2}_{\nu}$ = 1 for the edited data set eccentric orbit fit.
The edited data set excludes the points plotted as open
circles in the figure which
have large deviations from the best fit curve.
The orbital period is derived from the ASM light curve rather than
pulse timing. T0 corresponds to periastron passage for the eccentric
orbit fits and the phase of mid-eclipse for the circular fits.

\end{table}


\begin{figure}
\plotone{f1.eps}
\end{figure}


\begin{figure}
\plotone{f2.eps}
\end{figure}


\begin{figure}
\plotone{f3.eps}
\end{figure}


\begin{figure}
\plotone{f4.eps}
\end{figure}

\begin{figure}
\plotone{f5.eps}
\end{figure}

\begin{figure}
\plotone{f6.eps}
\end{figure}

\begin{figure}
\plotone{f7.eps}
\end{figure}


\begin{references}

\reference{}
Bildsten, L., et al., 1997, \apjs, 133, 367

\reference{}
Bradt, H.V., Rothschild, R.E., \& Swank, J.H., 1993, \aaps, 97, 355

\reference{}
Burke, B.E., Mountain, R.W., Daniels, P.J., Cooper, M.J., \&
Dolat, V.S., 1993, \procspie, 2006, 272

\reference{}
Clark, G.W., 2000, \apjl, 542, L131

\reference{}
Corbet, R.H.D., 1986, \mnras, 220, 1047

\reference{}
Corbet, R.H.D., Marshall, F.E., Peele, A.G., \& Takeshima, T.,
1999, \apj, 517, 956 (Paper I)

\reference{}
Gotthelf, E.V., Ueda, Y., Fujimoto, R., Kii, T., \& Yamaoka, K.,
2000, \apj, 543, 417

\reference{}
Hall, T.A., Finley, J.P., Corbet, R.H.D., \& Thomas, R.C., 2000,
\apj, 536, 450

\reference{}
Jahoda, K., Swank, J.H., Stark, M.J., Strohmayer, T., Zhang, W., \&
Morgan, E.H., 1996, ``EUV, X-ray and Gamma-ray Instrumentation for
Space Astronomy VII, O.H.W.  Siegmund \& M.A. Gummin, eds., SPIE 2808,
59, 1996

\reference{}
Jalota, L., Gotthelf, E.V., \&  Zoonematkermani, S., 1993,
\procspie, 1945, 453 

\reference{}
Levine, A.M., Bradt, N., Cui, W., Jernigan, J.G., Morgan, E.H.,
Remillard, R., Shirey, R.E., \& Smith, D.A., 1996, \apjl, 469, L33

\reference{}
Levine, A.M., Rappaport, S.A., \& Zojcheski, G., 2000,
\apj, 541, 194

\reference{}
Liu, Q.Z., van Paradijs, J., \& van den Heuvel, 2000,
\aap, 147, 25

\reference{}
Makishima, K., et al., 1996, \pasj, 48, 171

\reference{}
Ohashi, T., et al., 1996, \pasj, 48, 157

\reference{}
Remillard, R.A., \& Levine, A.M.,  1997, in Proceedings of ``All-sky
X-ray observations in the next decade'', p. 29, RIKEN, Japan, March,
1997.

\reference{}
Rothschild, R.E., Blanco, P.R., Gruber, D.E., Heindl, W.A., MacDonald,
D.R., Marsden, D.C., Pelling, M.R., Wayne, L.R., \& Hink, P.L., 1998,
\apj, 496, 538

\reference{}
Tanaka, Y., Inoue, H., \& Holt, S.S., 1994, \pasj, 46(3) L37

\end{references}
\end{document}